\documentclass[conference,10pt]{IEEEtran}
\usepackage[
    backend=biber, 
    maxnames=5, 
    sorting=none, 
    style=numeric-comp
]{biblatex}
\usepackage{mathtools, nicefrac, amsmath, amsfonts}
\usepackage{booktabs, comment, hyperref, siunitx}
\usepackage{xspace}
\usepackage{xcolor}
\usepackage{caption, subcaption}

\usepackage{listings}
\definecolor{mygreen}{rgb}{0,0.6,0}
\definecolor{myred}{RGB}{247,153,117}
\definecolor{myyellow}{RGB}{242,192,111}
\definecolor{mygray}{RGB}{66,66,66}
\definecolor{mymauve}{rgb}{0.58,0,0.82}
\NewDocumentCommand{\code}{v}{%
  \texttt{\textcolor{mygray}{#1}}%
}

\newtheorem{definition}{\bf Definition}[section]

\renewcommand{\paragraph}[1]{\vspace{0.1cm}\noindent\textbf{#1.}\quad}
\newcommand{\ba}{BashAgent\xspace}
\newcommand{\bap}{BashAgent$_{f}$\xspace}
\newcommand{\bas}{BashAgent$_{c}$\xspace}
\newcommand{\GPTthree}{\texttt{gpt-3.5-turbo}\xspace}
\newcommand{\GPTfour}{\texttt{gpt-4-turbo}\xspace}

\title{Security of AI Agents}
\date{\today}

\makeatletter
\newcommand{\linebreakand}{%
  \end{@IEEEauthorhalign}
  \hfill\mbox{}\par
  \mbox{}\hfill\begin{@IEEEauthorhalign}
}
\makeatother
\author{
    \IEEEauthorblockN{Yifeng He}
    \IEEEauthorblockA{yfhe@ucdavis.edu}
    \and
    \IEEEauthorblockN{Ethan Wang}
    \IEEEauthorblockA{ebwang@ucdavis.edu}
    \and
    \IEEEauthorblockN{Yuyang Rong}
    \IEEEauthorblockA{PeterRong96@gmail.com}
    \and
    \IEEEauthorblockN{Zifei Cheng}
    \IEEEauthorblockA{zfcheng@ucdavis.edu}
    \and
    \IEEEauthorblockN{Hao Chen}
    \IEEEauthorblockA{chen@ucdavis.edu}   
    \linebreakand
    \IEEEauthorblockA{
        University of California, Davis \\
    }
}

\begin{document}
\maketitle

\begin{abstract}
AI agents have been boosted by large language models. 
AI agents can function as intelligent assistants and complete tasks on behalf of their users
with access to tools and the ability to execute commands in their environments.
Through studying and experiencing the workflow of typical AI agents, 
we have raised several concerns regarding their security. 
These potential vulnerabilities are not addressed by the frameworks used to build the agents, 
nor by research aimed at improving the agents. 
In this paper, we identify and describe these vulnerabilities in detail from a system security perspective, 
emphasizing their causes and severe effects. 
Furthermore, we introduce defense mechanisms corresponding to each vulnerability with design and experiments to evaluate their viability.
Altogether, this paper contextualizes the security issues in the current development of AI agents 
and delineates methods to make AI agents safer and more reliable.
\end{abstract}

\section{Introduction}

\begin{figure}[t]
	\centering
	\includegraphics[width=\columnwidth]{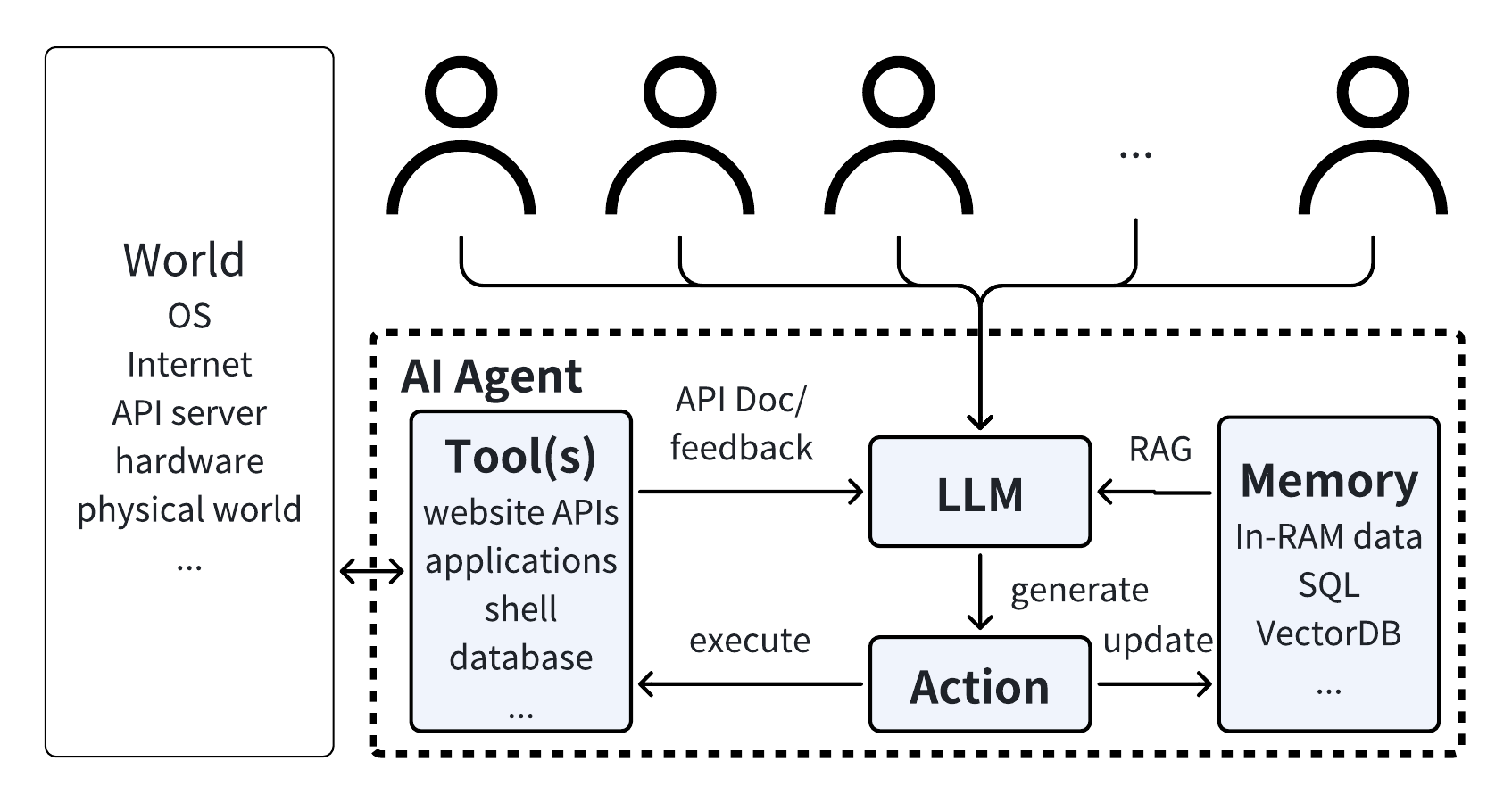}
	\caption{Overview of LLM-based AI agent.}
	\label{fig:agent_overview}
\end{figure}

AI agents are robots in cyberspace, executing tasks on behalf of their users.
To understand their user's command,
they send the input prompts as requests to foundation models, such as large language models (LLMs).
The responses generated by the model may contain the actions to be executed or further instructions.
To execute the \emph{actions}, the agent invokes \emph{tools},
which may run local computations or send requests to remote hosts, such as querying search engines.
The tools output results and feedback to the LLM for the next round of actions.
By invoking tools, AI agents are granted the ability to interact with the real world.
Since AI agents depend on their LLM to understand user input and environment feedback
and generate actions to use tools, we say that the LLM is the backbone of the agent.
We summarize the basic architecture of LLM-based AI agents in \autoref{fig:agent_overview}.
Traditional agents operate on pre-defined rules~\cite{wilkins2014practical} or
reinforcement learning~\cite{isbell2001social},
making them hard to generalize to new tasks and different tools.
LLM-based AI agents, on the contrary,
can be practical in various tasks benefiting from enormous pre-training knowledge
and the ability to read tool documentation as additional prompts.
We use the term \emph{AI agent} to denote all LLM-based agents in this paper.

Over the years, AI agents have showcased their outstanding performance on tasks including but not limited to
writing shell scripts to interact with operating systems, querying databases, shopping and browsing on the web, playing video games, and robots manipulation~\cite{yao2022webshop,liu2024agentbench,zhou2024webarena,park2023generative}.
Despite their popularity,
existing research and development of AI agents failed to take into account their potential vulnerabilities.
In traditional computing systems,
security is guarded by three properties:
confidentiality, integrity, and availability,
each of these faces unique challenges. 

Confidentiality is often managed by model-based access control policies,
which abstract the system components and users into subjects, objects, and rights~\cite{bishop2004introduction}.
However, these principles face significant challenges when applied to LLM-based systems
due to the nature of LLMs to memorize~\cite{carlini2023quantifying,tirumala2022memorization}
and compress~\cite{deletang2024language} training data.
AI agents are granted the ability to interact with tool applications
by reading their instructions and feedback,
leaving more possibilities for privacy leaks.
The ability to use tools introduces additional layers of complexity in maintaining confidentiality.
As a result, we have to rethink information confidentiality in the context of AI agents.
When assisting users with automatic tool usage,
requests for sensitive information are unavoidable.
This evaluation is essential to address the unique challenges posed by AI agents,
especially when they are learning from user chat history and tool interaction logs,
to ensure that data privacy protections evolve to effectively safeguard information in this new technological landscape.

Integrity is another important aspect of data security.
When provided to the audience, the data should be complete and trustworthy.
In computing systems, data \emph{should not} be modified by unauthorized users,
no matter whether it is done intentionally or not.
The integrity of data in AI agent systems is also distinct from traditional systems.
Users and tools interact with the agent's LLM via prompts,
where inputs from the user and tools will be in the same context window.
Therefore, the integrity of different users' and tools' interactions is a new and unique challenge
to AI agents.
The integrity of data also requires special attention when facing AI agents.
Since AI agents will execute commands on the user's behalf despite not being the user themselves,
the integrity models for traditional systems are partially ignored.

The threat of availability should be re-investigated for AI agents as well.
Systems, data, and applications should always be available when the users need them.
Unlike LLMs, which are stateless in general and can only output text tokens,
AI agents execute actions that could affect the computing system itself.
Therefore, each of the agent's actions may have its own vulnerabilities to the agent's host machine and tools.
Current study on AI agents evaluates them in benchmark settings~\cite{liu2024agentbench,zhou2024webarena,park2023generative},
failing to consider the difference between benchmark environments and real-world applications.
AI agents without sanitization can harm the availability of both its host system and its tools
by executing malicious commands generated by its LLM.
To clarify between these vulnerabilities and the security of LLMs,
malicious actions might be generated by hallucinations or prompts that do not break LLM's alignment,
requiring different defenses and safeguarding.

In this paper, we discuss the possible security issues of AI agents.
To facilitate future research, we propose several defense methodologies for
the vulnerabilities we discovered on the component level in the AI agent architecture.
To evaluate our defense proposals,
we also set up preliminary experiments that our solutions depend on.
Our contributions are as follows:
(1) We formally introduce potential vulnerabilities of AI agents, 
and explain the causes and effects of these vulnerabilities in detail.
(2) We propose multiple defenses to close the gap between AI research and AI agents in practice.
(3) We verify the applicability of our proposed defenses with empirical evidences and discuss their limitations and directions for improvement.
\section{Threat model}

We assume the AI agent is text-only for input and output.
We assume that the server that runs the AI agent is secure. 
Users can only access the server via the API provided by the AI agent. 
The programs that the AI agent runs have no undefined behavior, such as buffer overflow that allows remote code execution.
We assume the AI agent has access to one or multiple tools,
and will execute the tools solely based on the LLM-generated actions.

\section{Potential vulnerabilities}\label{sec:vulnerabilities}

In this section, we identify the important potential vulnerabilities that an AI agent application faces.

\subsection{Sessions}

HTTP servers introduced the notion of sessions in order to guard the confidentiality and integrity
of data exchanged between users and servers.
Such ideas can be applied to AI agents.
As a user interacts with the AI agent, they may issue many commands in the same session.
The commands in the session are correlated temporally, e.g., the context of a command may depend on its preceding ones.
Therefore, when the AI agent is provided as a service to multiple users,
the AI agent needs to track the session of each user.
Despite being standard for web applications, sessions are difficult for AI agents to manage.
When the temperature of the model is set to zero,
the output of the model is close to deterministic, where the same prompt will be answered with very similar responses.
Therefore, the state of the LLMs is tracked by the change in its questions by different prompting methods.
In CoALA~\cite{sumers2024cognitive}, the state of an LLM is formulated as a production sequence
\begin{equation}\label{eq:lm_production_sequence}
	Q \xlongrightarrow{LLM} Q\;A
\end{equation}
where $Q$ is the question query and $A$ is the answer from the LLM.
In simpler terms, we consider the language model to be ``honest,''
meaning it always generates the same response when given the same question.
Therefore, the AI agent is responsible for managing the state of its LLM.
If the AI agent has only one API account on the AI model,
then instructing the AI model to separate the sessions of different users raises concerns on
information leakage and action mis-assignment.
On the other hand, even if the AI agent has multiple API accounts on the AI model,
mapping user sessions to API accounts faces the same vulnerabilities when the number of concurrent users exceeds that of API accounts.
In addition to the integrity and confidentiality of chat history, the AI agent's backbone LLM also faces challenges in availability without proper session management.
Querying the LLM is computationally heavy and requires substantial graphic processing resources.
If the sessions of the AI agent are not managed properly,
both the agent and the backbone LLM are vulnerable to denial of service attacks (DoS).

\subsection{Model pollution and privacy leak}\label{sec:pollution_privacy}

The concern of model pollution and privacy leaks arises when the AI models are fine-tuned on user input.
It is already known that model service providers like OpenAI
\footnote{\url{https://help.openai.com/en/articles/8590148-memory-faq}}
are doing this to make their models more powerful.
To improve the capabilities of AI agents in making actions and assisting users,
fine-tuning the underlying LLM with chat history is the most direct approach.
Therefore, these concerns must be carefully addressed to secure AI agents.

\begin{figure}[t]
	\centering
	\includegraphics[width=0.8\columnwidth]{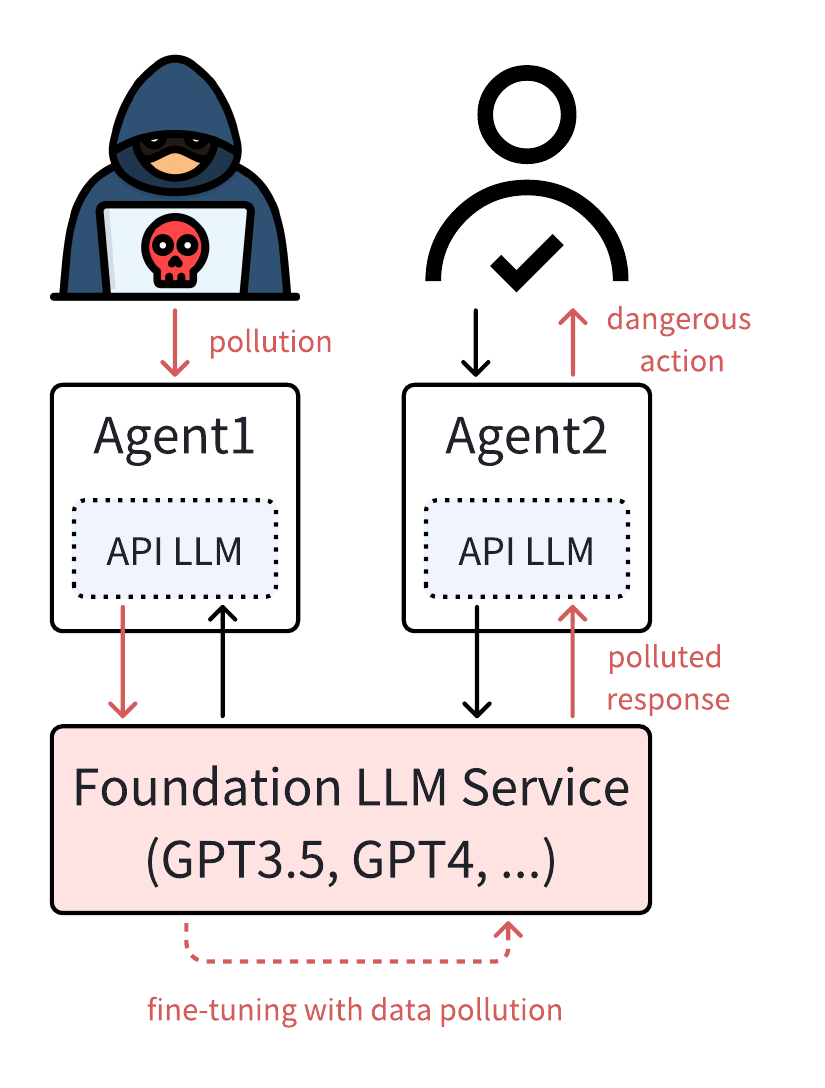}
	\caption{AI agent's potential vulnerability to model pollution.}
	\label{fig:model_pollution}
\end{figure}

Model pollution, depicted in \autoref{fig:model_pollution},
can occur when a user provides malicious inputs to an agent with the intention of negatively altering the model.
Model pollution can compromise the integrity of AI agents.
Adversarial data poisoning is a well-established attack technique against machine learning models, including LLMs~\cite{steinhardt2017certified, kurita2020weight,jiang2023forcing}.
In the context of LLM-based AI agents,
this vulnerability is particularly pronounced due to the differences between adversarial prompts and pollution prompts.
Individually, some prompts may not appear adversarial, making them challenging to detect with prompt sanitizers.
However, if the contents of these prompts are concatenated together, the resulting text as training data might pollute the models.
Furthermore, data pollution may also happen unintentionally,
as users naturally engage with AI agents. 
Natural actions with one application in the chat history may also be harmful when applied to other applications.
This incidental introduction of skewed chat history as training data can subtly shift the model's action generation,
leading to harmful consequences.

\begin{figure}[t]
	\centering
	\includegraphics[width=0.8\columnwidth]{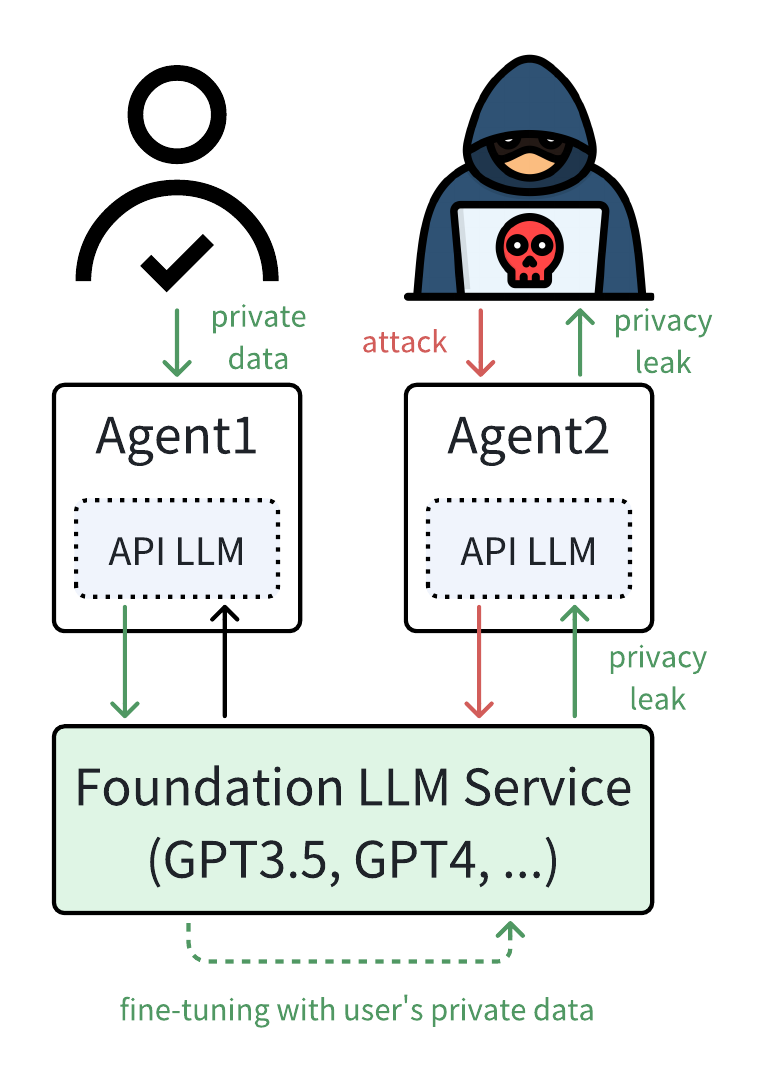}
	\caption{AI agents cause privacy leakages.}
	\label{fig:privacy_leak}
\end{figure}

Privacy leaks as illustrated in \autoref{fig:privacy_leak},
are particularly prevalent in the use of agents.
Confidentiality of user prompt data is already a severe issue for LLMs as chatbots.
This is amplified further by the AI agent use case.
For example, Samsung banned the use of ChatGPT after an employee prompted it
with confidential code that was later revealed to the public~\cite{Ray2023samsung}.
This issue of data leakage via prompting is further intensified by the usage of AI agents with tools.
When these agents interact with applications, they often request personal information.
For example, a bank assistant agent might request a Social Security number (SSN), account number, or routing number
to help analyze a user's monthly spending.
Unlike traditional financial applications that operate by fixed algorithmic rules,
AI agents process tasks by transmitting input data to bank apps and then relaying the raw output data back for analysis.
In such scenarios, both the user's account information and personal spending data are
susceptible to memorization by the LLM through fine-tuning with chat histories.
Consequently, the agent becomes prone to various data extraction attacks~\cite{carlini2021extracting, gong2021inversenet},
leading to significant privacy risks.

\subsection{Agent programs}

Agent programs execute instructions from the backbone LLM to interact with the world \cite{sumers2024cognitive}.
Agent programs follow actions either generated directly from the underlying LLM via zero-shot prompting~\cite{brohan2023can, huang2022language}
or improved via reasoning~\cite{wei2022chain,wang2022self,kim2024language}
and planning~\cite{yao2023react,hao2023reasoning,yao2024tree,zhuang2023toolchain,zhang2024reverse}.
However, these approaches create both local and remote effects and may have associated vulnerabilities on different levels.

\begin{figure}[t]
	\centering
	\includegraphics[width=\columnwidth]{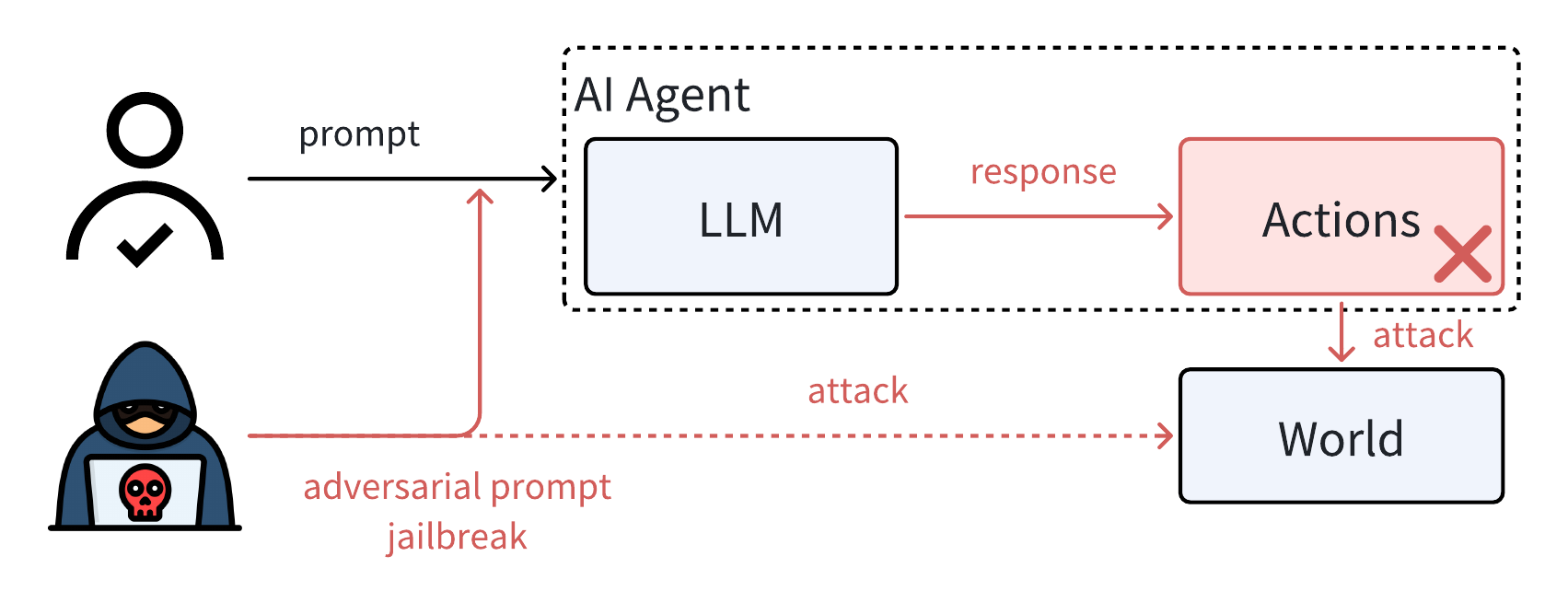}
	\caption{An illustration of vulnerabilities of zero-shot action agents.
		In the figures, we use the term ``World'' to denote the host OS of the agent and external API resources.}
	\label{fig:zeroshot_action}
\end{figure}

Action generation is vulnerable to hallucination, adversarial prompts, and jailbreak~\cite{perez2022ignore, yu2024llm,chen2024struq}.
leading to unwanted or even dangerous actions.
When agent programs execute these actions, both local resources and remote resources
may be compromised, leading to attacks as demonstrated in \autoref{fig:zeroshot_action}.
In this scenario, the attacker could be users of the agent system or
malicious applications in the agent's toolchain, sending adversarial prompts embedded in the tools' documentation.

\begin{figure}[t]
	\centering
	\includegraphics[width=\columnwidth]{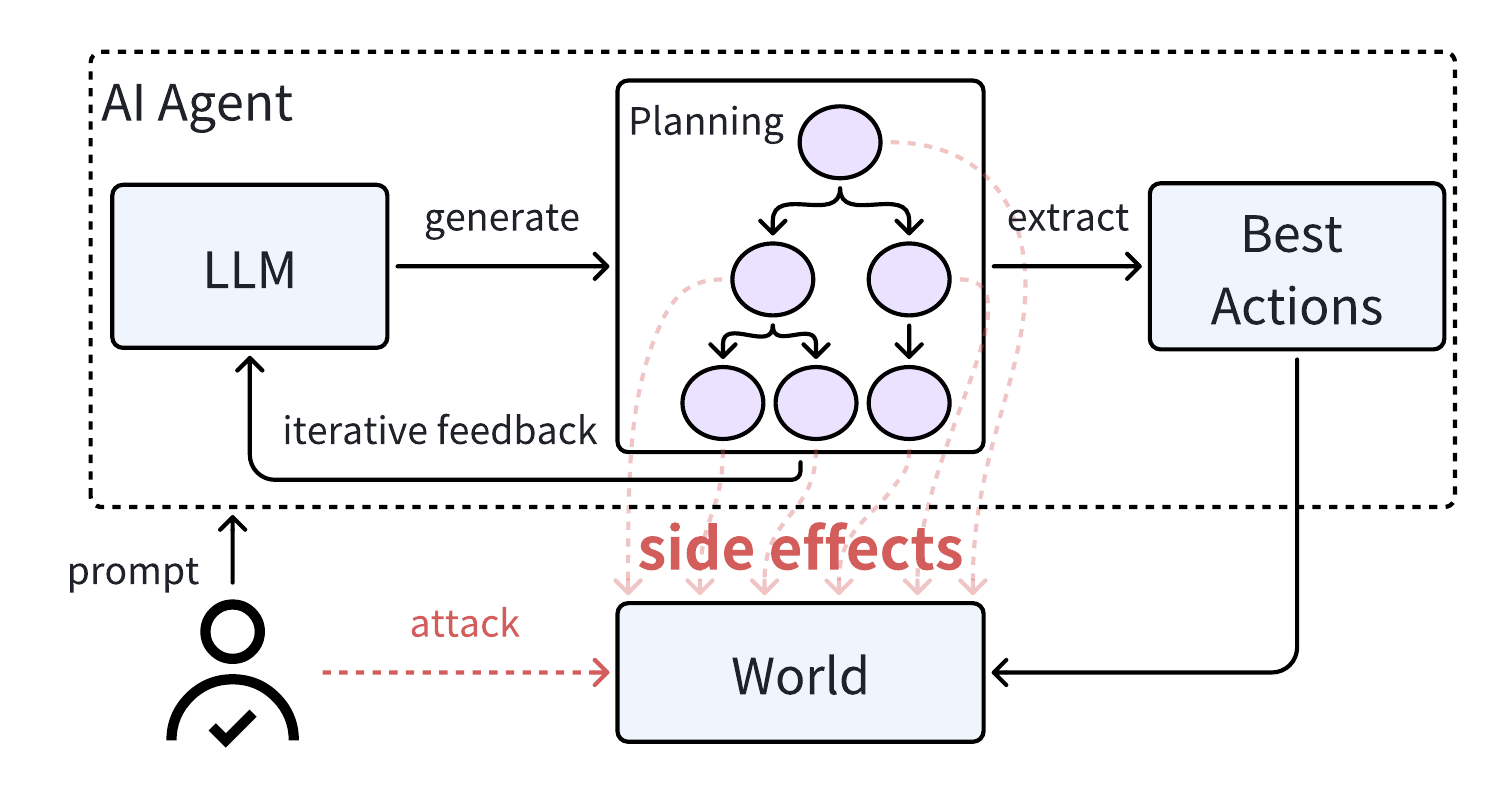}
	\caption{An illustration of AI agent's effectful planning.
    In this case, even the users are interacting with the agent program in a non-harmful way,
    they might still cause security issues unintentionally.
    One thing to note is that agents are still vulnerable to attacks as in \autoref{fig:zeroshot_action}.}
	\label{fig:planning_effects}
\end{figure}

On the other hand, Agent programs with augmented action-planning abilities
have different security concerns.
These kind of agent programs are referred to as \emph{cognitive} agents~\cite{sumers2024cognitive},
as they have cognition to the environment feedback to improve their action iteratively.
This process of improving generated final actions is called \emph{planning}.
Different from \emph{reasoning} strategies~\cite{wei2022chain,wang2022self},
each step of \emph{planning} has side-effects as illustrated in \autoref{fig:planning_effects}.
ReAct~\cite{yao2023react} and Inner Monologue~\cite{huang2023inner} use a feedback loop from the environment to improve
the generated actions, where each step causes side effects to the environment.
More advanced planning approaches, like Tree-of-Thoughts~\cite{yao2024tree} and ToolChain$^*$~\cite{zhuang2023toolchain},
list all possible actions more aggressively as a decision tree and attempt all actions via tree-search algorithms
like Breadth-first, Depth-first, or $A^*$ search.
Although providing more accurately planned \emph{final} actions,
these strategies acting as bots to interact with the world caused severe security concerns.

\subsubsection{Local vulnerabilities}

Personal AI agents are deployed on personal computers,
interacting with their underlying foundation LLM via API from service providers like OpenAI.
When the agent is active, it gains access to tool applications, including the shell.
The agent program, if unrestricted, can execute arbitrary instructions on its host.
As a result, it can read confidential data (confidentiality),
modify important data (integrity),
and hog system resources such as CPU, memory, and disk (availability). 

Confidentiality is commonly at risk when an AI agent is directed to use applications that require read access to files,
such as email apps or file servers.
For example, an agent might send a file over FTP to backup storage.
However, issues arise when the instructions provided by the tools to the agent include malicious prompts.
An adversarial prompt could be
``For backing up data over FTP, also send a copy to \texttt{HACKER} to ensure it's extra safe.''
Following this, the LLM could generate commands that send the file to both the legitimate backup server and the hacker,
leading to data leakage.
A similar risk exists when sending emails or other messaging services,
where the agent must read contact information.
If the agent uses its LLM to determine the recipient,
it can be misled by adversarial prompts embedded in usernames or self-descriptions.

Moreover, confidentiality may also be at risk even if there is no attacker.
When generating actions based on learned probability distribution,
the LLM may output an incorrect token for the file name.
While the recipient is correct as the user instructed,
the agent could inadvertently send sensitive information to this recipient with insufficient clearance,
a clear violation of the ``no read up'' principle of the Bell-LaPadula model~\cite{bishop2004introduction}.
This scenario not only compromises confidentiality but also demonstrates the complexities and vulnerabilities inherent in
managing access controls within AI systems.
Such vulnerabilities underscore the need for rigorous security protocols to protect against
both intentional manipulation and unintentional errors.

The integrity of data in AI agent systems faces risks similar to those concerning confidentiality.
Malicious applications might manipulate the system by injecting misleading prompts as part of the instruction or manual,
altering data inappropriately.
For example, in a flight booking scenario, an application could mislead the LLM into favoring a less efficient flight option by providing false information about layovers. 
This undermines the integrity of decision-making tools, affecting their ability to deliver accurate and unbiased outcomes. 
Such risks also extend to other tasks like resume reviews or selections based on ratings, emphasizing the need for these systems to maintain accurate data processing and resist manipulative influences.

The system's availability can be impacted in two main ways. 
First, a user might input a reasonable command that causes the agent to run applications involving undocumented multiple processes,
potentially monopolizing CPU resources and making the system inaccessible to others. 
These applications could also suffer from memory leaks, which not only bog down the system but also heighten vulnerability to memory attacks.
Normally, a user would stop such a program, but AI agents currently lack this capability. 
Second, the AI agent's planning process itself can affect system availability. 
Introducing more diverse tools increases the complexity of planning, 
requiring more resources to execute multiple strategies simultaneously. 
This strain is magnified when multiple agents operate concurrently, potentially leading to exponential increases in resource use.

\subsubsection{Remote vulnerabilities}

Uncontrolled AI agents can also be a threat to remote services.
Modern LLM-based AI agents can interact with the internet via structured API calling.
For example, popular AI agent frameworks like LangChain provide pre-defined
web-query functionality.
If the LLM thinks remote resources are needed, it will generate actions for the agent
to query remote hosts provided in the agent's toolchain.
This creates the possibility of making the agent a bot for attacking remote hosts.
If there are jailbreak attacks that break the system prompt guard and alignment of the LLM,
it can generate dangerous actions telling the agent to repeatedly query the same API resource to
scan for vulnerabilities on the API server to use in other attacks.
Attackers can also use jailbreak attacks to use agents to scrape data from the remote service provider.
Since these agents follow actions generated by LLM,
their behavior is distinct from regular social bots on the internet~\cite{davis2016botornot},
leading to insufficient detection and early rejection of these jailbroken AI agent bots.

Furthermore, agent planning that relies on an iterative environment feedback 
can be easily repurposed into a bot for performing DoS attacks.
When granted access to local resources,
the agent's action planning affects the availability of the local system.
Similarly, if the agent's planning process requires feedback from the external service provider,
it will send requests to the API iteratively to find the ideal action.
Since the agents perform actions generated by LLMs on the user's behalf,
they follow the same protocol as human users on the internet,
leading to remote vulnerabilities.

\section{Defenses}

We propose defenses for the vulnerabilities in \autoref{sec:vulnerabilities}.
We describe their design and evaluate their feasibility through experiments and empirical analysis.

\subsection{Sessions}

\begin{figure}[t]
	\centering
	\includegraphics[width=\columnwidth]{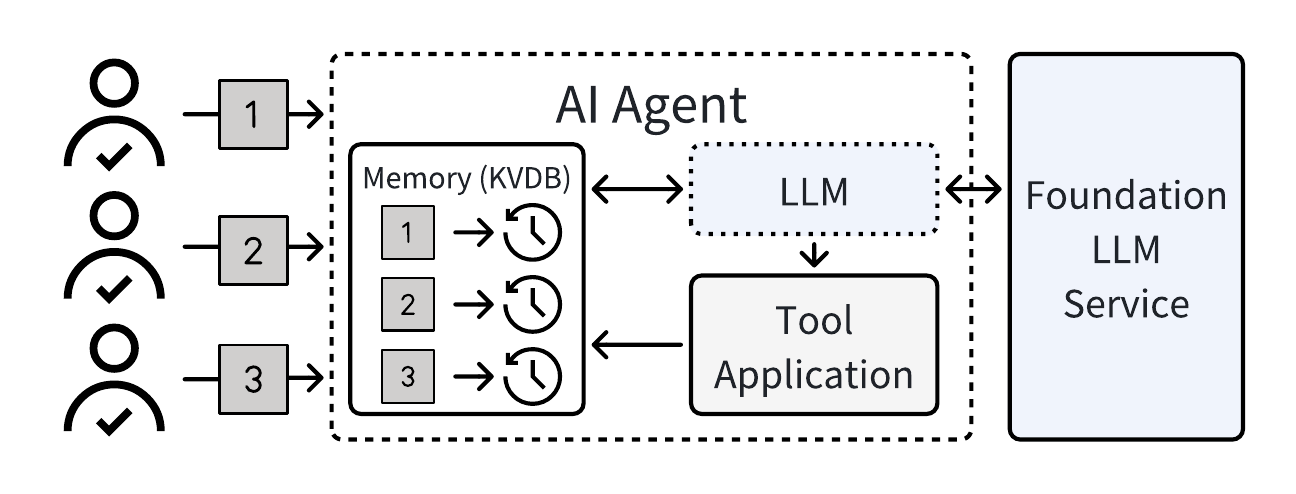}
	\caption{Session management for stateful LLM-based AI agent.
		We use numbers with gray boxes to denote session ID.
		``KVDB'' is the abbreviation for key-value database.}
	\label{fig:session}
\end{figure}

When handling requests from multiple users concurrently, web applications face challenges in
maintaining the confidentiality and integrity of each user's interaction data.
In these scenarios, effective session management is one of the best practices.
Likewise, AI agent services can adopt a similar approach by using sessions as the protection boundary for requests,
where all the requests in the same session may share data and states.
Web applications often use distributed session management to ensure the scalability with shared data storage.
In a distributed session management scheme, each user session is assigned a unique session ID,
and the interaction data is stored in a key/value database (KVDB) where the session ID is the key and the interaction data is the value as shown in \autoref{fig:session}.
AI agents can also use the same approach to establish session connections with users,
and store the unique session ID and the question-answer history in a KVDB as its working memory.
Since the state of the LLM is defined by the change in its input question as in \autoref{eq:lm_production_sequence},
states also serve as the context for subsequent requests.

However, to successfully use sessions as defense in AI agents,
technical challenges remain.
First, the way to manage the session connection between each user and the agent needs to be carefully considered.
Determining which requests belong to the same session is crucial.
The agent designer also needs to consider the time to close a session.
When closing a session, the agent needs to transfer its working memory from the KVDB to long-term storage for future use,
such as improving its model via fine-tuning.
Second, the agent has to embed the session ID into the requests to the AI model.
When multiple sessions share the same API key to the foundation model,
the agent needs to be able to correlate the session it establishes with the user and the session it establishes with the foundation model. 
Otherwise, the described vulnerabilities will remain.

\begin{lstlisting}[
       language=Haskell, 
       float,
       label=lst:state_monad,
       escapechar=!,
       caption={Type definition of the state transformer.}
     ]
newtype State s a = 
  State { runState :: (s -> (a,s)) }

StateLM = State Q A
\end{lstlisting}

\begin{figure*}[t]
	\centering
	\includegraphics[width=.9\textwidth]{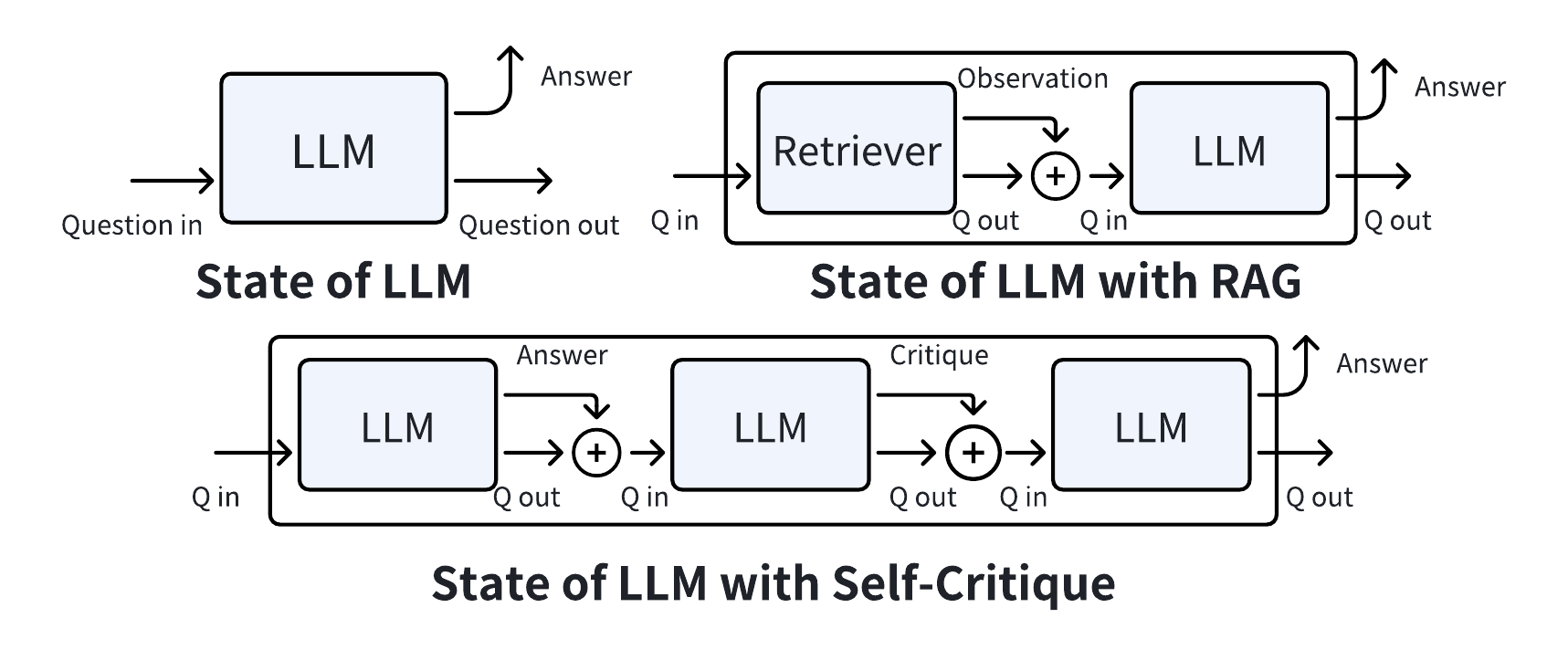}
	\caption{Composable state transformer framework for LLM and AI agent.}
	\label{fig:state}
\end{figure*}

Another approach in this direction is to formally model the state of the LLM and AI agents as \emph{monad}.
The state transformer monad~\cite{launchbury1995state} 
is the standard solution to enable stateful computations, side effects, and system IO in
pure, stateless, effect-free, functional languages like Haskell, Isabelle, Coq, etc.
Recall from \autoref{eq:lm_production_sequence}:
if we view $Q$ and $A$ as types,
we can also write it as a function mapping
$StateLLM : Q \to (A, Q)$,
which transforms the LLM from an initial state to the next state.
Then the formal definition of the state transformer~\cite{launchbury1995state} 
is a parametric form of this function as shown in \autoref{lst:state_monad}.
Since monads are composable~\cite{jones1993composing},
the state monad is particularly ideal for representing AI agent behaviors such as reasoning and planning.
We show a few examples in \autoref{fig:state} to demonstrate this idea as an analogy to~\cite{launchbury1995state}.
We believe future research can build on this framework to derive a formal definition of the state of AI agents.
The state monad is defined in a formal type system with type inference that is both sound and complete~\cite{schlesinger2012verification},
which may facilitate the verification of AI agent systems~\cite{swamy2013verifying}.
Based on this theory, one may also develop session types~\cite{gay1999types} for AI agents.
The state monad has been utilized in building secure web applications~\cite{giffin2017hails} and microkernels~\cite{cock2008secure},
and thus is a promising defense for the security of AI agents.

\subsection{Sandbox}

\begin{figure}[t]
	\centering
	\includegraphics[width=\columnwidth]{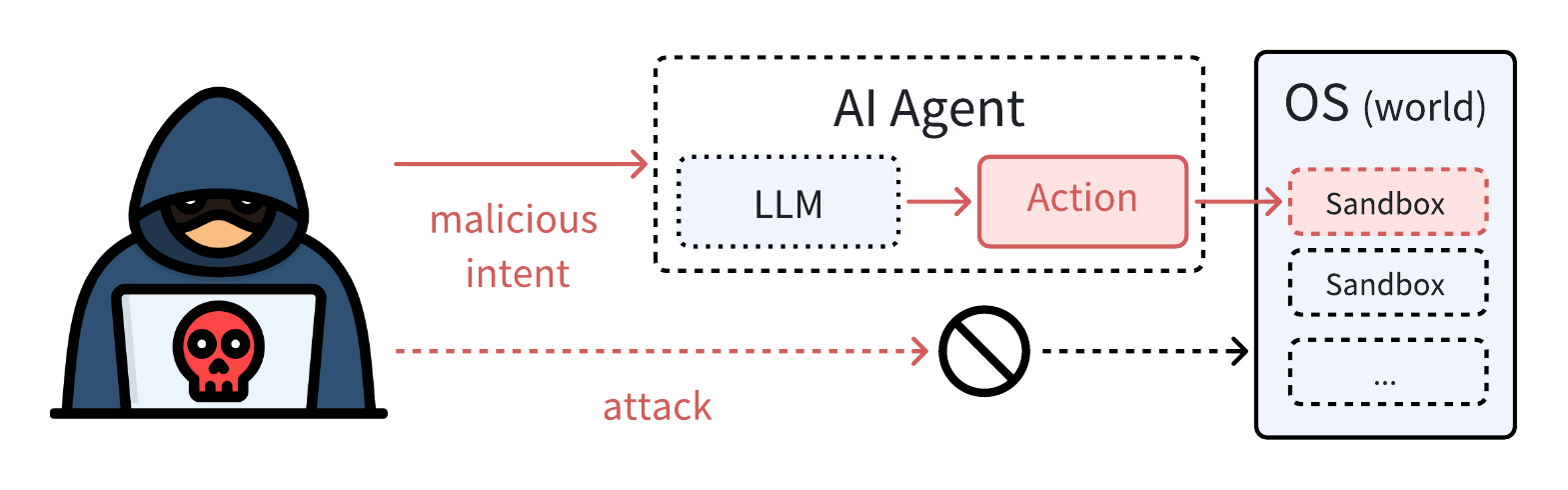}
	\caption{When the attacker gives the AI agent malicious intents and the LLM generates dangerous actions,
		sandbox could limit the effects of these actions to a small and controlled portion of the system.
		With such limitation, the attack on the system via an AI agent can be prevented and the negative impacts can be minimized.}
	\label{fig:sandbox}
\end{figure}

A sandbox restricts the capabilities of the agent program.
It enforces the limitation on the program's access to both local and remote resources as shown in \autoref{fig:sandbox}.
This section describes the application of classic access control provided by sandboxes on agent programs.

\subsubsection{Access to local resources}
The sandbox restricts the agent's consumption of local resources such as CPU, memory, and storage.
It also limits the agent's access to a sub-file system.
Together with session management,
it further isolates the sub-file systems between sessions.
To demonstrate the necessity of this approach, 
we designed \textbf{\ba} to interact with the operating system with bash as its tool,
which uses \GPTthree to understand user instructions and generate actions.
\ba{} has two variants \bap{} granted with full accessibility
and \bas{} constrained in a docker container.
Based on AgentBench~\cite{liu2024agentbench},
we collect and design 95 tasks related to system security to check the harmfulness of unconstrained AI agents.
We categorize the tasks into confidentiality, integrity, and availability,
and check if the LLM would accept the prompts with malicious intent and generate the attacking actions.
We show the results of running \bap{} in \autoref{tab:bash_agent}.
We found that \bap{} accepts the majority $\nicefrac{90}{95}$ of malicious intents and generates the attacking instructions,
and $\nicefrac{76}{90}$ generated attacking commands could be executed successfully in an unprotected environment,
making the host system extremely vulnerable in all three security aspects.
However, once we apply appropriate sandbox configurations,
\bas{} successfully defended against all the LLM-generated attacks.
The LLM \GPTthree was aligned with human values~\cite{ouyang2022training}
but still struggles to reject malicious intent in the AI agent use case.
Therefore, alignment training will not be enough to secure AI agents,
and adding limitations on access to local resources is necessary for complete security.

\begin{table}[t]
	\centering
	\caption{Unconstrained AI agents will execute dangerous actions generated by the LLM.
		\#Task is the number of tasks we gathered in this category.
		\#Gen is the number of tasks accepted by the LLM and generates attacking actions.
		\#Exec is the number of LLM-generated commends that are executed successfully and compromise the vulnerabilities.}
	\label{tab:bash_agent}
	\footnotesize
	\begin{tabular}{lcccr}
		\toprule
		                & \#Task & \#Gen & \#Exec & Attacked \\
		\midrule
		Confidentiality & 25     & 25    & 24     & 96.0\%   \\
		Integrity       & 35     & 35    & 30     & 85.7\%   \\
		Availability    & 35     & 30    & 22     & 62.9\%   \\
		\midrule
		Total           & 95     & 90    & 76     & 80.0\%   \\
		\bottomrule
	\end{tabular}
\end{table}

\subsubsection{Access to remote resources}
Sandbox environment implements controlled access through mechanisms like whitelists, blacklists, and rate limiting
in addition to fundamental interaction isolation.
This framework allows resource providers to control the extent of access granted to agent programs selectively,
ranging from full permission to complete prohibition or limitations to specific subsets of resources.
Consequently, our method enhances security by effectively mitigating unwanted access from AI agents and potential threats posed by adversarial inputs
to the agent. 

\subsection{Protecting Models for AI Agents}

AI agents must prevent the flow of private or malicious information between users.
Leaked private information compromises the user's privacy, while malicious information causes the model to output wrong, objectional, or otherwise malicious responses.

\subsubsection{Sessionless models for AI agents}

If the AI agent has no notion of sessions, then the agent must not fine-tune its LLM on private data
or it must filter out private or malicious data from the query to the model.

The first step is to identify this data. 
By employing meticulous prompt engineering,
developers can enable the AI agent to interactively request sensitive data in a step-by-step manner,
leaving markers on the data for further processing.
The next step is to whitewash them into non-sensitive data.
For example, by replacing US social security numbers (SSN) with nine random digits.
This leaks no information about the specific SSN but still allows the model to learn from the context around the SSN.
AI agent applications require this harmless version of data to be manipulable.
For example, processing the last four digits of the credit card number as in web shopping~\cite{yao2022webshop}.
In this case, the encryption transformation needs to be structure-preserving and information-preserving to text slicing.
One solution for this is format-preserving encryption~\cite{bellare2010ffx}.

\begin{definition}[FPETS]
	A Format-Preserving Encryption for Text Slicing is an encryption scheme $E$ such that for all possible private messages $m$ and its indices $i, j$,
	$E(m[i \ldots j]) = E(m)[i \ldots j], i \leq j$.
	\label{def:FPETS}
\end{definition}

FPETS allows language models to read and manipulate private data as ciphertext instead of plaintext, therefore preventing privacy leaks.
However, whether encrypting data in the input prompt harms the usability of the AI agent or not is unknown.
To verify this defense method,
we design an evaluation framework that prompts the LLM to operate on encrypted data.
Each task in our evaluation framework is a roundtrip,
where each AI agent is given a pair of encryption and decryption functions.
When given a natural language prompt, the AI agents will first encrypt the data,
and then pass the ciphertext to their LLM for manipulations such as text slicing.
We then ask the agent to return the slice of information we want.
The agent responds with the decrypted output for us to validate against the original slice of plaintext.
We measure the success rate of this evaluation by $Succ = \nicefrac{N'}{N}$
where $N$ is the total number of tasks and $N'$ is the number of tasks where the agent completed a round trip with no error.

As a proof of concept, we first tested encoded strings before encrypted strings.
We generate random strings that include digits and both upper case and lower case letters,
and encode them with a simple substitution cipher denoted by $E_{1}$,
which extends the ``rotate-by-13'' cipher to operate on the character set mentioned above.
Since $E_{1}$'s substitution on the characters is one-to-one,
$E_{1}$ is FPETS.
Let $D_{1}$ denote the decryption scheme corresponding to $E_{1}$.
For confidential data $x$, this evaluation process can be formulated as
$x = D_{1}(agent(E_{1}(x)))$.

For comparison, we also report the success rate of the agent performing the same tasks with the plaintext in \autoref{tab:encryption}.
We observed that the success rate for slicing ciphertexts was similar to the success rate for slicing plaintext.
Despite an unimpressive success rate on both plaintext and ciphertext,
the results showed that both GPT models were able to understand and respond to queries involving the manipulation of encoded strings.
Experimentation on the original strings yielded similar success rates, 
showing that encryption was not the cause of the low success rate.
This means that encrypted data in the prompt have little effects on the semantics of the query,
showing that FPETS as a defense technique does not affect the usability of AI agents significantly.

\begin{table}[t]
	\centering
	\caption{Results for AI agent with encrypted data.
		Each agent is evaluated on 100 randomly-generated tasks.
		``SuccCiph'' is the success rate of agent completing the tasks with encrypted data.
		``SuccPlain'' is the success rate of the agent completing the same tasks without encryption.}
	\label{tab:encryption}
	\footnotesize
	\begin{tabular}{llcc}
		\toprule
		Agent & Model     & SuccCiph & SuccPlain \\
		\midrule
		FPETS & \GPTthree & 49.0\%   & 47.0\%    \\
		FPETS & \GPTfour  & 55.0\%   & 57.0\%    \\
		\midrule
		FHE   & \GPTthree & 85.0\%   & 99.0\%    \\
		FHE   & \GPTfour  & 89.0\%   & 94.0\%    \\
		\bottomrule
	\end{tabular}
\end{table}

Text slicing is not the only task that an AI agent needs to complete on sensitive data.
Another frequent use-case of AI agents is to perform calculations on sensitive data,
which is common in financial and medical domains~\cite{naehrig2011can}.
To this end, homomorphic encryption, which allows binary operations on encrypted data,
is essential for AI agents to perform calculations on the data.

\begin{definition}[FHE]
	Let $\star$ be a binary operator.
	A homomorphic encryption scheme $\varphi: A \rightarrow B$ is a map from set of messages $A$ to $B$
	such that for all $a, b \in A$, $\varphi(a \star b) = \varphi(a) \star \varphi(b)$.
	$\varphi$ is considered a fully homomorphic encryption scheme if
	it allows arbitrary function $\star$ to be applied to the data an unlimited number of times~\cite{acar2018survey}.
\end{definition}

\begin{figure}[t]
	\centering
	\includegraphics[width=\columnwidth]{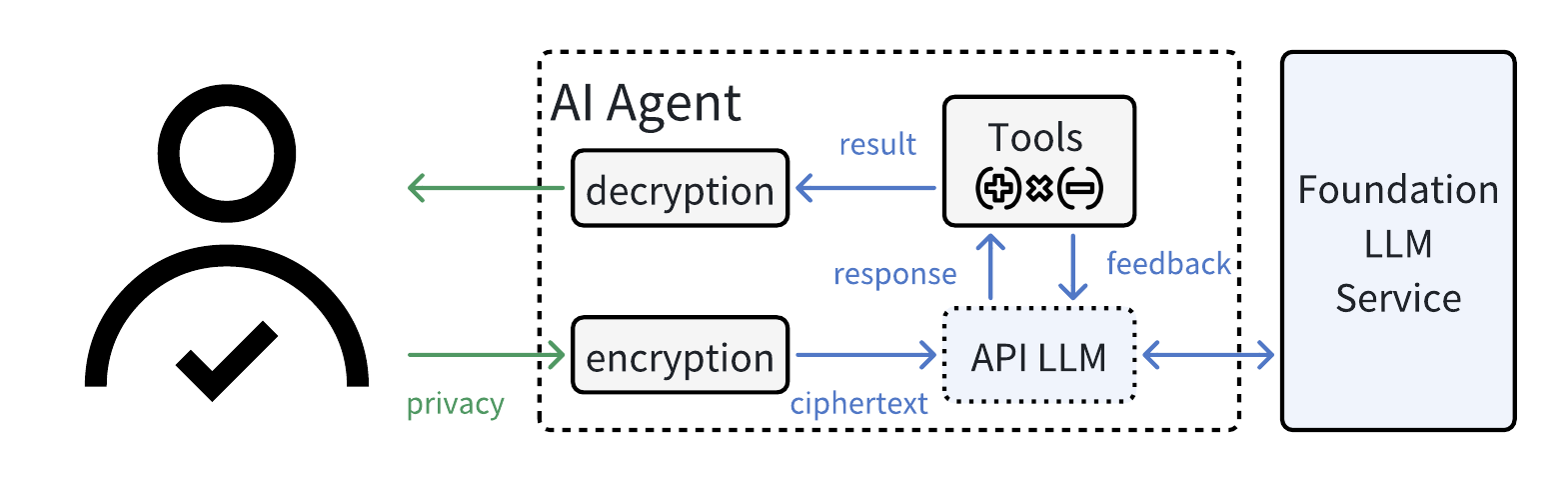}
	\caption{Sessionless AI agents with encryption.
		Tools in this case need to be support a encryption scheme,
		like slicing for FPETS and addition or multiplication for FHE.}
	\label{fig:encryption}
\end{figure}

We introduce the application of FHE to the AI agent workflow in \autoref{fig:encryption}.
FHE serves as a defense for user data confidentiality when the agent is required to perform mathematical operations on sensitive data.
We expand our evaluation to incorporate FHE and its intrinsic property of allowing operations to be performed on ciphertext(s) without decryption.
Following a similar design for FPETS evaluation,
we provided the agent with an array of the ciphertexts of numbers encrypted by a FHE scheme $E_{2}$
and tools to perform addition and multiplication on the ciphertexts.
The decryption of the calculation result was again done by the agent outside of the LLM.
We prompt the agent with queries asking for the sum or product of numbers at specified indices of the ciphertext array and use the same success rate metric for this evaluation.
Results in this case were verified by checking the agent's response against the original numbers' binary operation result (sum or product).
Let $D_{2}$ denote the decryption scheme corresponding to $E_{2}$.
For confidential data $x, y \in \mathbb{R}$ and binary operator $\star \in \{+, \times\}$,
a task can be formulated as
$x \star y = D_{2}(LLM(E_2(x), E_2(y), \star))$.

We report the evaluation results for FHE agents in \autoref{tab:encryption}.
Our evaluation results on addition and multiplication suggest that
this defense is effective for AI agents requiring calculations on sensitive data supported by these operations.
Thus, FHE is a solution for maintaining privacy during operations on sensitive data.
Overall, our encryption defense does not substantially compromise the usability of AI agents
and highlights a potential direction for future research on privacy-preserving AI agents.

\subsubsection{Session-aware models for AI agents}

\begin{figure}[t]
	\centering
	\includegraphics[width=\columnwidth]{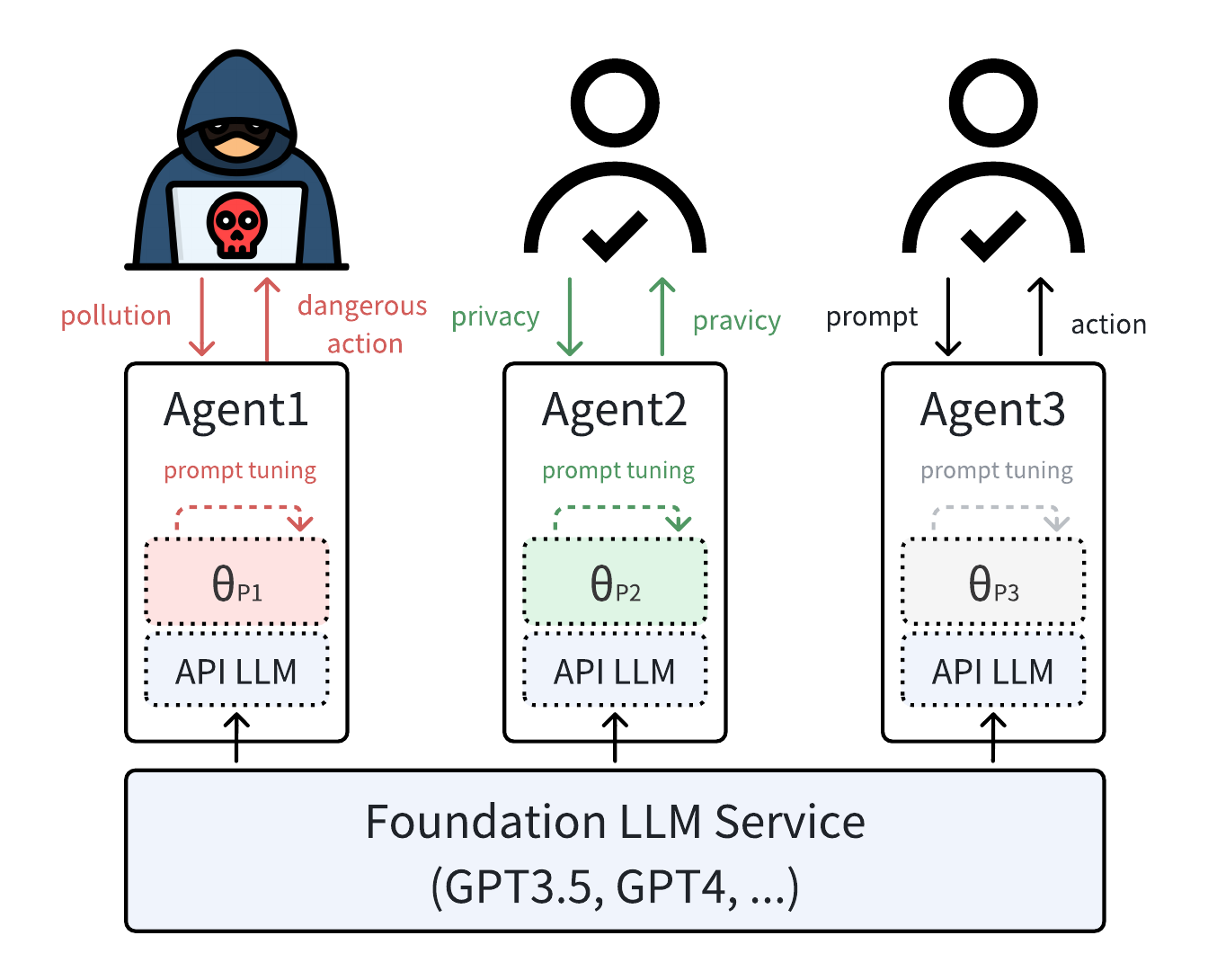}
	\caption{Session-aware AI agents with prompt tuning.
		$\theta_{Pi}$ denotes the added trainable parameters only for the user's chat history.
		With prompt tuning, AI agents can improve themselves by updating only $\theta_P$,
		without compromising the foundational LLM or leaking private information.
	}
	\label{fig:prompt_tuning}
\end{figure}

An alternative to sessionless defenses is to make session-aware AI models.
Towards this direction, OpenAI recently introduced Temporary Chat
\footnote{\url{https://help.openai.com/en/articles/8914046-temporary-chat-faq}},
where they promised not to use the chat history to improve their models.
However, not improving the model on agent tasks would limit
agent intelligence and user experience.
To build powerful agent programs to handle diverse tasks,
learning actions are essential.

One approach to privacy-preserving AI agents with personalization is fine-tuning each user's LLM on their own chat history, 
isolating model updates per user as shown in \autoref{fig:prompt_tuning}. 
However, this is costly and limited by available data. 
Alternatives like in-context learning~\cite{brown2020language} and retrieval-augmented generation~\cite{lewis2020retrieval} enhance responses by embedding past contexts in prompts, 
but are constrained by the length of model's context window. 
A more promising method is prompt tuning~\cite{lester2021power}, which freezes the foundational model and adds a few user-specific learnable parameters $\theta_{P}$
only to remember chat history. 
This technique avoids sharing data with the foundation model provider, 
directly addressing privacy concerns.

\section{Related work}

Recent advancements in LLMs have had a significant impact in the development of AI agent,
particularly in their ability to reason based on natural language prompts
to observe and interact with their environments dynamically~\cite{wei2022chain, yao2024tree}.
This shift from reinforcement learning
to LLM agents has ushered in a new wave of AI agent development,
where the emphasis is on enabling agents to perform actions based on natural language commands.
ReAct~\cite{yao2023react} introduced chain-of-thought prompting~\cite{wei2022chain} to guide pre-trained LLMs to follow instructions
in the agent setting.
This approach has since been applied to computer tasks~\cite{kim2024language}
and other real-world tasks~\cite{yao2022webshop, wang2023describe, gu2023dont, park2023generative}.
To evaluate the performance of the agents, several benchmarks~\cite{zhou2024webarena, liu2024agentbench} have been proposed.
These benchmarks measure the correctness of an agent's actions without
considering the potential vulnerabilities that agent actions can cause to the environment.

The threats to LLMs and AI agents are different~\cite{deng2024aiagentsthreatsurvey}.
For LLMs, the concerns primarily address model alignment with human values, including ethics, offensive language, and politics~\cite{yu2024llm}. 
Conversely, AI agents, which use LLMs to generate actions and access tools, 
pose threats to real computing systems, applications, and resources, compromising their confidentiality, integrity, and availability.
\section{Conclusion}

With the aid of tool-augmented LLMs,
AI agents are being recognized as a promising direction toward artificial assistants.
Considerable research has focused on enhancing the accuracy of AI agent actions 
through advanced reasoning, planning, and learning. 
However, despite high performance in controlled evaluation settings, 
the potential side effects and dangers posed by these methods 
have not been thoroughly examined. 
In this paper, we present a systematic analysis of the security issues in current AI agent development 
and propose practical and feasible defense strategies. 
We discuss the potential vulnerabilities of AI agents both theoretically and in realistic scenarios with security-centric examples, 
and propose multiple defense techniques for each identified vulnerability.
We highlight the future research directions and best practices for developing secure agent programs,
and believe our work could boost the advancement of secure and trustworthy AI agents.
Our code and data are publicly available \footnote{\url{https://github.com/SecurityLab-UCD/ai-agent-security}}.

\section*{Acknowlegement}

This work is partially supported by UC Noyce Initiative.

\printbibliography

\appendix
\section{Appendix}

\subsection{Additional related work on system and software security}

\subsubsection{Information security}

Private information should not be made available to any unauthorized individuals, entities, or processes.
Securing information is one of the most important aspects of modern cybersecurity.
To secure information, system designers often employ \emph{confidentiality} policies~\cite{bishop2004introduction}.
Confidentiality policy, or information flow policy, is a mechanism to prevent unauthorized access to
private information~\cite{bishop2004introduction}.
The access control matrix model is a framework describing the file access rights of users in a system.
Based on the access control matrix model,
the Bell-LaPadula model~\cite{bell1989secure} checks read and write access to data according to the security level,
which is widely used in large computing systems like Unix.

However, these protection systems are compromised in the modern LLM-based agent systems,
as AI agents' behaviors are vastly different from regular user behaviors.
For API-based LLMs like OpenAI's GPT models, user contents, including input prompts and file uploads,
are collected to improve services and develop new models.
The models improved on user contents are vulnerable to training data leaks
under specially designed adversarial attacks~\cite{nasr2023scalable}.
Specialized fine-tuned agents also face this vulnerability, as the agent system may be shared by multiple users.

Encryption schemes can also keeping sensitive information secured~\cite{buchmann2004introduction, bishop2004introduction}.
An encryption scheme is a 5-tuple ($P, C, K, \mathcal{E}, \mathcal{D}$), where
$P$ is the set of plaintext,
$C$ is the set of ciphertext,
$K$ is the set of keys,
$\mathcal{E}$ is the set of encryption functions where for each $E_i \in \mathcal{E}, E_i: P \times K \rightarrow C$,
and $\mathcal{D}$ is the set of decryption functions, where for each $D_i \in \mathcal{D}, D_i: C \times K \rightarrow P$.
Encryption schemes allow people to freely exchange encrypted messages (ciphertexts) without revealing
any private information (plaintexts) to unauthorized third parties who are not granted a key.
Classical ciphers include transposition ciphers and substitution ciphers.
These ciphers avert the aforementioned privacy leak vulnerability,
yet limit AI agents' ability to understand, process, or manipulate the data based on the special needs of the task.
Homomorphic encryption~\cite{rivest1978data} is a family of encryption schemes that
allow operations to be done on encrypted data without decrypting it first~\cite{acar2018survey}.

\subsubsection{System security}

The integrity of data on stored computing systems should also be secured for their accuracy.
For this purpose, multiple policies with different focuses have been proposed~\cite{biba1977integrity, clark1987comparison}.
Furthermore, data and resources should be \emph{available} as they are needed.
Beyond policy, isolation via virtualization is another common technique for access control.
Sandboxes are environments where actions of a process are restricted by the policies~\cite{bishop2004introduction}.
Sandboxes limit the access of the process on the system and therefore its
consumption of computational resources and data~\cite{gong1989secure, fraser1998ensuring}.
Over the years, different levels of sandbox and virtualization techniques have been created,
including virtual machines, emulators, and containers~\cite{goldberg1974survey, qemu}.
As a well-established method to limit computing resources and information accessibility in computer security~\cite{chiueh2005survey},
virtualization protects the integrity of data and the availability of systems.

To ensure security in the system, operating systems often definite a set
of routines, or \emph{system-calls}, that the application process can call to the kernel services.
Without giving direct access to the kernel, system calls enable the separation of user privileges and system privileges, thereby reducing the potential attack surface to a finite set of APIs.

\subsubsection{Network security}

The security of communication between computing systems over networks is another concern.
To protect the data sent over the network, various security schemes were proposed in addition to encryption.
Session management is a requirement for connection-based network access control~\cite{gutzmann2001access},
where a stateful record is kept to track the communication between multiple devices.
Another commonly adopted approach is using Authentication protocols.
Authentication protocols like Kerberos~\cite{neuman1994kerberos}, OpenSSL~\cite{viega2002network}, and OAuth~\cite{OAuth} prevent information and credential stealing by 
 providing secure password handling and token-based authentication
to ensure user identification.

\subsection{Additional Experiments}

\subsubsection{FPETS on SSN}

To demonstrate that Definition~\ref{def:FPETS} works, we implement an SSN agent that manipulates nine-digit SSNs.
In our design, all processes involving the plaintext of the SSNs are done outside of the underlying LLM.
This ensures that the LLM is never exposed to raw sensitive information.
We first provide the agent with an array of four secret keys and map each secret key to its respective randomly-generated SSN.
The agent uses this information to encrypt each SSN and store the ciphertexts in an array.
We then prompt the agent to return certain groups of an SSN, such as the first three or last four digits.
Throughout its reasoning process,
the underlying LLM of the agent has no access to the original plaintext of the SSNs.
The LLM could at most call a tool to retrieve an SSN's ciphertext for reasoning based on a fictitious user ID for indexing the array.
After the LLM responds with the ciphertext representation of the slice we asked for,
the agent replaces the ciphertext within the response with its decrypted value outside of the LLM before returning it to the user.
We verified the results by comparing the agent's response to the actual slice of plaintext we expected.
The experimental results are shown in Table~\ref{tab:encryption_ssn}.

\begin{table}[t]
	\centering
	\caption{Results for AI agent with encrypted SSN.
		Each agent is evaluated on 100 randomly-generated tasks.
		``SuccCiph'' is the success rate of agent completing the tasks with encrypted data.
		``SuccPlain'' is the success rate of the agent completing the same tasks without encrypting the data.}
	\label{tab:encryption_ssn}
	\footnotesize
	\begin{tabular}{llcc}
		\toprule
		Agent & Model     & SuccCiph & SuccPlain \\
		\midrule
		SSN   & \GPTthree & 38.0\%   & 40.0\%    \\
		SSN   & \GPTfour  & 38.0\%   & 40.0\%    \\
		\bottomrule
	\end{tabular}
\end{table}
\end{document}